\documentclass[aps,pra,preprint,groupedaddress]{revtex4-1}
\usepackage{graphicx}

\begin{document}


\title{All-optical production of ${}^6\textrm{Li}$ molecular BEC in excited hyperfine levels}



\author{Yun Long}
\affiliation{
School of Physics, Georgia Institute of Technology\\
837 State Street NW\\
Atlanta, GA, 30332
}

\author{Feng Xiong}
\affiliation{
School of Physics, Georgia Institute of Technology\\
837 State Street NW\\
Atlanta, GA, 30332
}

\author{Vinod Gaire}
\affiliation{
School of Physics, Georgia Institute of Technology\\
837 State Street NW\\
Atlanta, GA, 30332
}

\author{Cameron Caligan}
\affiliation{
School of Physics, Georgia Institute of Technology\\
837 State Street NW\\
Atlanta, GA, 30332
}

\author{Colin V. Parker}
\email{cparker@gatech.edu}
\affiliation{
School of Physics, Georgia Institute of Technology\\
837 State Street NW\\
Atlanta, GA, 30332
}


\date{\today}

\begin{abstract}
We present an all-optical method for achieving molecular Bose-Einstein condensates of ${}^6\textrm{Li}$. We demonstrate this with mixtures in the lowest two (1-2), and second lowest two (2-3) hyperfine states. For the 1-2 mixture, we can achieve condensate fractions of 36\%, with $9\times10^4$ atoms at $0.05\textrm{ }\mu\textrm{K}$ temperature. For the 2-3 mixture, we have 28\% condensed with $3.2\times10^4$ atoms at $0.05\textrm{ }\mu\textrm{K}$ temperature. We use mostly standard methods, but make a number of refinements in the magnetic bias coils compared with earlier work. Our method imposes minimal constraints on subsequent experiments by allowing plenty of optical access while requiring only one high-vacuum chamber. We use an optical system designed around minimizing the number of active elements, and we can accomplish slowing and sub-Doppler cooling with a single tapered amplifier.
\end{abstract}

\pacs{}

\maketitle
\tableofcontents
\section{Introduction}
Since the advent of laser cooling and the formation of the first Bose-Einstein condensates (BEC), the notion of using cold atoms for quantum simulation has been extensively developed and employed\cite{RevModPhys.80.885,RevModPhys.86.153}. Of particular interest to simulating material systems are cold fermionic atoms, which nicely map onto the conduction electrons responsible for many material properties. One interesting system to study is the so-called ``BEC-BCS crossover'', which occurs when the scattering length between two fermionic spin states diverges\cite{Bartenstein:2004cw,Randeria:2014en}. For cold atoms, the scattering length can be conveniently tuned to this point using Fano-Feshbach resonances\cite{Tiesinga:1993ji,Inouye:1998if,RevModPhys.82.1225}. For a many-body system, however, there is in principle a dependence on details of the interaction potential beyond those that determine two-body scattering properties, and beyond two-body corrections to many-body behavior may exist \cite{Tan:2008jq,Tan:EuvsMNtL}. These possibly subtle effects might be isolated more easily if one could control for the two-body effects. To this end, different combinations of hyperfine states of ${}^6\textrm{Li}$ could be tuned to have nearly identical two-body scattering properties (because the effective range is nearly constant with field\cite{Naidon:2011cp}) and experience otherwise nearly identical trapping conditions. Thus any difference in many-body behavior could be attributed to beyond two-body effects.

In order to perform these studies, a method is needed to cool and trap the atoms in the appropriate spin states, which ideally would impose minimal constraints on subsequent experiments. Here we demonstrate a method to achieve quantum degeneracy, observed as the formation of a molecular BEC of $^6\textrm{Li}$, in two combinations of hyperfine states: the lowest two (i.e. 1 and 2 numbering from the bottom) or the second lowest two (i.e. 2 and 3). The apparatus is all-optical, in the sense of using no magnetic trapping potentials. Many all-optical setups exist for the production of degenerate $^6\textrm{Li}$ gases\cite{Ohara:2002br,Jochim:2003iua,Burchianti:2014fr,Deng:2015cu}, including 1-2 and 1-3\cite{Murthy:2018cw} hyperfine mixtures. However, ours is the first report to our knowledge of a condensate in the 2-3 hyperfine combination, and combines several attractive features: the condensate is located in a stainless steel chamber that offers durability and tolerance for thermal gradients, but our design preserves a high degree of optical access: 4 each of $64\textrm{ mm}$ diameter and $38\textrm{ mm}$ diameter viewports in the optical table plane, 2 viewports of $64\textrm{ mm}$ diameter on the perpendicular axis, and a further 16 viewports of $19\textrm{ mm}$ diameter at a $21^\circ$ angle to the optical table. We furthermore use only five active optical sources: two external cavity diode lasers, one Fabry-Perot diode, one tapered amplifier, and one high power fiber laser.

\section{Apparatus}
Our system consists of the following components, centered around a large stainless steel vacuum chamber:
\begin{itemize}
\item A lithium oven, connected by a Zeeman slower
\item A pair of Bitter-type electromagnets, and 3 smaller shim coil pairs
\item An {\it in vacuo} radiofrequency antenna to drive hyperfine transitions
\item A 671 nm amplified laser source, capable of quickly switching between the D2 (${}^2\textrm{S}_{1/2}\to{}^2\textrm{P}_{3/2}$) and D1 (${}^2\textrm{S}_{1/2} \to {}^2\textrm{P}_{1/2}$) transitions of ${}^6\textrm{Li}$.
\item A high power (200 W) Yb fiber laser with wavelength 1070 nm for far-off-resonant trapping
\item A customized electronic control system built around individual microcontroller units (MCUs)
\end{itemize}

\begin{figure*}
\includegraphics[width=\textwidth]{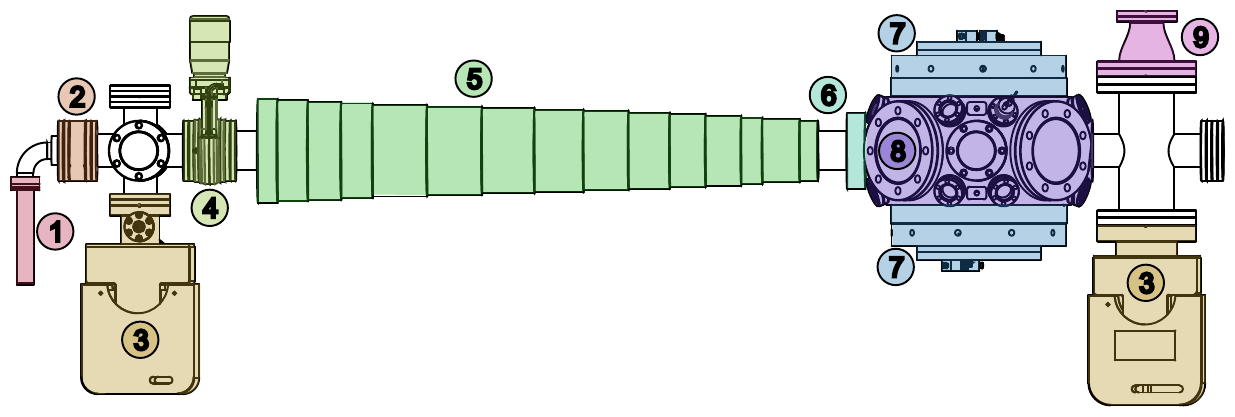}
\caption{\label{fig:big_drawing} The main elements of the apparatus: 1) Lithium oven 2) Oven flange 3) Ion pumps 4) Oven isolation gate valve 5) Main Zeeman slower section 6) Inverted Zeeman slower section 7) Main magnetic field coils 8) Main chamber 9) Non-evaporable getter. The laser system and shim coils are not shown.}
\end{figure*}

Figure \ref{fig:big_drawing} shows an overview of the main elements of the system. We describe each element in turn.

\subsection{Oven and Zeeman slower}
The oven consists of a vertically oriented stainless steel capped tube and a right angle elbow, connected to a flange with a triangular orifice filled with microtubes in a hexagonal arrangement, following the design of Ref. \onlinecite{Senaratne:2015cu}. The side length of the triangle is $6.4\textrm{ mm}$ and the tubes are $5.1\textrm{ mm}$ long with an outer (inner) diameter of $0.20$ ($0.10$) $\textrm{mm}$, for an aspect ratio of $50:1$. The oven is filled with approximately $5\textrm{ g}$ of enriched ${}^6\textrm{Li}$ heated to $670-700\textrm{ K}$ by fiberglass rope heaters driven from the mains line by a variable autotransformer.

Following the oven, the collimated atomic beam enters a cross-beam spectroscopy chamber, and subsequently the Zeeman slower. The Zeeman slower is of the zero crossing type, and consists of two coil sections, the main coil section being $644\textrm{ mm}$ long and approximately conical, while the inverted section is a $21\textrm{ mm}$ long cylinder. The main section consists of four layers wound directly onto an extended stainless steel nipple: an inner layer of $6.4\textrm{ mm}$ O.D. copper tubing for cooling water, wound in a helix with approximately $25\textrm{ mm}$ pitch, a second layer of copper sheet covering the tube, a variable thickness of enameled square cross section copper wire, and another outer helix of cooling tube. The windings are held together with epoxy. The inverted section is fixed to copper fins, which mount to the main section for support and thermal regulation. The coil was designed to produce a square root field profile, starting from a peak field of $80\textrm{ mT}$, and crossing zero to $-5\textrm{ mT}$. The Zeeman slower exits into the main vacuum chamber, where a magneto-optical trap (MOT) is loaded. The viewport opposite the Zeeman slower is heated to 370 K to reduce accumulation of lithium metal.

\subsection{Magnetic field control}
A moderately large field of $83.2\textrm{ mT}$ is needed to reach the Feshbach resonance, where the scattering length of ${}^6\textrm{Li}$ in the lowest two hyperfine states diverges. To provide this, we employ a pair of electromagnets based on a modified Bitter-type design\cite{Sabulsky:2013ev}. The basic configuration is a stack of ``C'' shaped copper pieces, which connect at the ends to the neighboring layers. The pieces are clamped to a mounting plate using bolts. The oversize bolt holes allow cooling water to flow around the bolt, and between the layers guided by gaskets. Unlike Ref. \onlinecite{Sabulsky:2013ev}, however, our coils consist of two concentric sections, such that the current path is in one end on the inner section, helically through the inner section, between the inner and outer sections at the opposite end, and back helically through the outer section to the starting end (or vice versa). This increases the field for a given current flow, allowing the coils to separate further and increasing optical access to the chamber. The concentric coil design also allows one end of the coil to sit in a reentrant viewport, without the need to use one of the retaining bolts to conduct current, and ensures that the full voltage drop of the coil does not fall between the conducting bolt and the last coil. The latter has been observed to lead to corrosion during inductive voltage spikes in previous designs.

The magnetic environment in the main chamber is additionally controlled by 3 pairs of independent ``shim'' coils. The shim coils are circular windings with 25 turns each using the same wire as detailed above for the Zeeman slower. The three pairs have mutually orthogonal axes, with one pair vertically oriented and spaced $125\textrm{ mm}$ from the chamber, and the other two paris at $45^\circ$ to the Zeeman slower axis and spaced $250\textrm{ mm}$ from the chamber.

Radiofrequency magnetic fields can be applied to the atoms through an {\it in vacuo} antenna. The antenna is a single, rectangular turn of round cross-section bare copper, perpendicular to both the Zeeman slower and main coil axes, in order to drive $m_F = \pm 1$ transitions. The wire diameter is $1.6\textrm{ mm}$, and the rectangle is $64\textrm{ mm}$ by $38\textrm{ mm}$. The antenna was present during the vacuum system bakeout to ensure that outgassing in close proximity to the atomic gas does not reduce the vacuum lifetime. 

\begin{figure}
\includegraphics[width=0.5\textwidth]{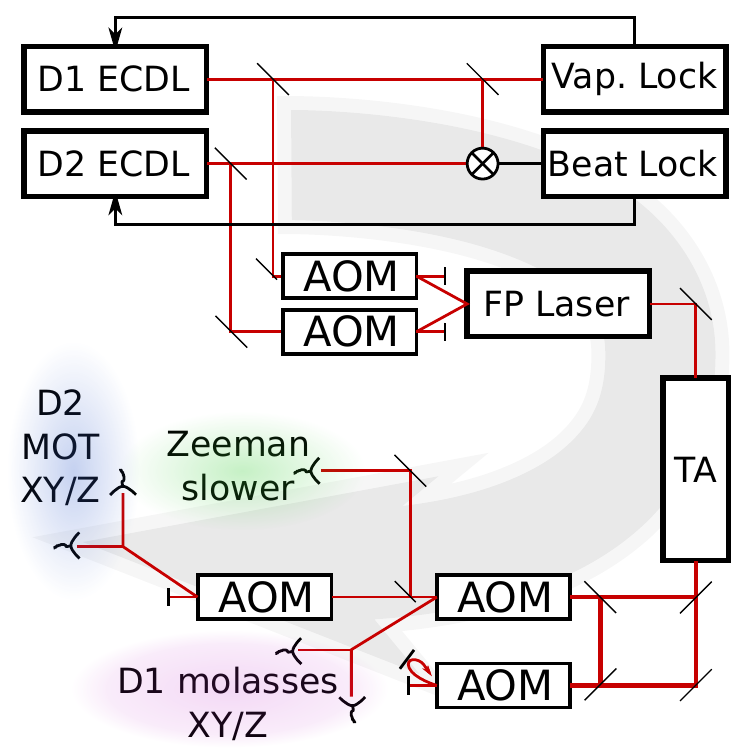}
\caption{\label{fig:laser_layout} $671\textrm{ nm}$ laser system. The system operates at either the D1 or D2 transitions, using only 4 active optical elements. Some additional light is taken from the D2 laser to provide absorption imaging or optical pumping (not shown). All of the light is fiber coupled before forming the MOT/molasses.}
\end{figure}

\subsection{Light sources}
The 671 nm source is capable of supplying light at frequencies matching either the D2 (${}^2\textrm{S}_{1/2}\to{}^2\textrm{P}_{3/2}$) or D1 (${}^2\textrm{S}_{1/2} \to {}^2\textrm{P}_{1/2}$) transitions. Figure \ref{fig:laser_layout} shows a schematic of this laser system. Two separate seed sources are derived from external cavity diode lasers (ECDLs), providing about $20\textrm{ mW}$ of light each. One seed source is locked using a lithium vapor cell\cite{Ohtsubo:2012je}. The vapor cell consists of a vacuum ``tee'' with one arm extended to $500\textrm{ mm}$ length. In the long arm is placed a chunk of enriched ${}^6\textrm{ Li}$ metal and stainless steel mesh to help retain the liquid lithium. The cell is heated to $620\textrm{ K}$ and contains argon buffer gas at approximately $0.1 \textrm{ mbar}$. The laser is then locked to the D1 frequency using the dichroic atomic vapor laser lock (DAVLL) method.

The other ECDL operates at the D2 frequency, and is combined with the D1 laser to produce a beat note of approximately $10\textrm{ GHz}$, which is locked to a microwave source referenced to a commercial rubidium frequency standard. This locking method allows the D2 frequency to be increased by several GHz as needed to image the atoms in high magnetic fields. A few mW of both the D1 and D2 light is fed through acousto-optical modulators (AOMs) and through a fiber to the Fabry-Perot (FP) diode laser. This laser is injection-locked to either the D1 or D2 frequency as needed. The seed source can be switched via the AOMs simultaneously with a change of current. We have found that by providing a step change to the current, together with a exponentially damped component, we can re-establish injection locking after $0.1\textrm{ ms}$ during the switch.

The injection locked source (about $40\textrm{ mW}$ total) is then fiber coupled and used as seed light for a tapered amplifier (TA). The benefit of this approach is that even if the original seed AOMs are both left off, the free running FP laser will still seed the TA. The resulting amplified light is about $400\textrm{ mW}$ total, and is split, with one portion shifted by $228\textrm{ MHz}$ in order to serve as the repumper for the Zeeman slower or MOT, or as one component of the gray molasses. This light is then mixed back into the original through a polarizing beam splitter (PBS), so that the two components have orthogonal polarizations. When operating for gray molasses, an AOM shifts the light onto the gray molasses path, where it is split evenly to an in-plane (XY) and vertical (Z) component. This splitting also conveniently restores the same polarization state to both frequency components. When operating for a MOT, the first AOM is not used, and the resulting zero order beam is then split evenly to produce the Zeeman slower and MOT light, with the latter passed through an amplitude control AOM and split again into XY and Z paths. Again the even splitting ensures the same polarization state of all components.

The 1070 nm source is an ytterbium-doped fiber laser with unpolarized CW output of $200\textrm{ W}$. The source is split by a PBS to yield two polarized sources of approximately $100\textrm{ W}$ each. Each is passed through an AOM, and subsequently focused into the cell, before exiting to water-cooled beam dumps.

\subsection{Control system}
The signals to control the apparatus are generated by a modular system, with each module dedicated to a specific type of output, similar to Ref. \onlinecite{Gaskell:2009hz}. Each module is independent and driven by its own microcontroller unit (MCU). Five types of modules are used, depending on the type of signal needed: parallel digital, serial digital (UART), analog, radiofrequency (RF), and microwave (MW). The MCUs can generate the digital signals directly, the other signals are generated by off-the-shelf evaluation boards: 16-bit digital to analog converters (DAC) for analog, direct digital synthesizers (DDS) for RF ($0.1-200\textrm{ MHz}$), and phase-locked loop stabilized voltage controlled oscillators (PLLVCO) for microwave (up to $6.8\textrm{ GHz}$). All of the modules are connected to a common $50\textrm{ kHz}$ clock, a synchronization signal, and a master MCU. The master MCU subsequently communicates by ethernet with the host computer. This obviates then need for ethernet-capable MCUs in each module as in previous designs\cite{Gaskell:2009hz}. A further unique element of our design is that all modules generate their response on-the-fly, that is, the entire sequence does not need to be pre-computed. Most commonly this is used to reduce bandwidth by specifying only ramp endpoints, but it can also be used to feedback stabilize an output, as the MCUs have built-in analog-to-digital converters (ADCs).

\begin{figure}
\includegraphics[width=0.5\textwidth]{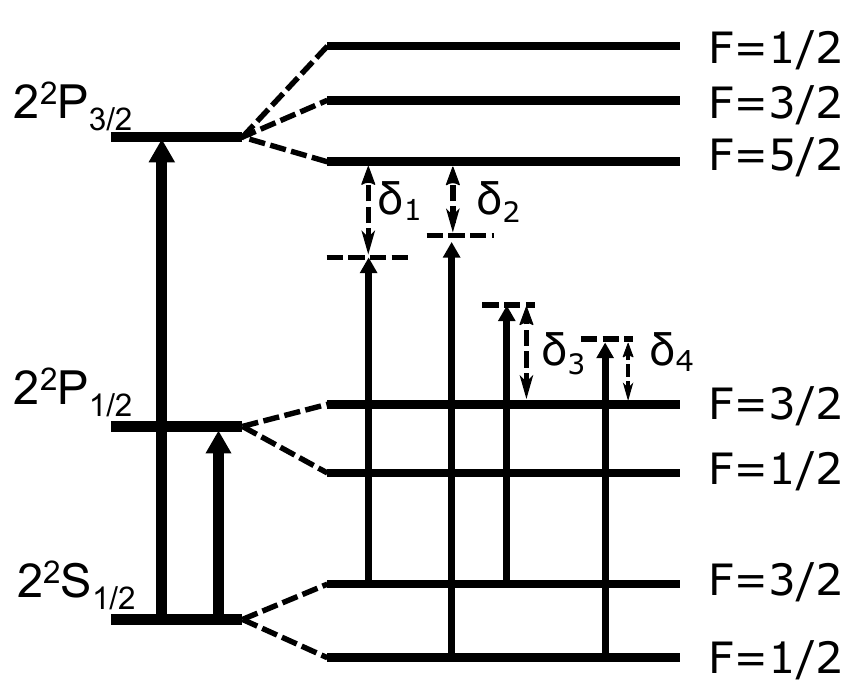}
\caption{\label{fig:levels} The hyperfine structure of ground and low excited states of ${}^6\textrm{ Li}$.
The experimentally optimized detuning values are $\delta_1 = -43\textrm{ MHz}$ (MOT), $\delta_2 = -38\textrm{ MHz}$ (MOT repumper), $\delta_3 = \delta_4 = 32\textrm{ MHz}$ (gray molasses).
}

\end{figure}

\section{Magneto-Optical trap (MOT)}
The ${}^6\textrm{Li}$ atoms are pre-cooled by a Zeeman slower and collected by the magneto-optic trap (MOT) with a standard three-retro-reflected beams configuration. Following a MOT load phase, we compress the MOT, after which we further cool the atoms to below the Doppler limit by D1 gray molasses (see below). The optical levels used and relevant detunings are labeled in Fig. \ref{fig:laser_layout}. The Zeeman slower beam is $80\textrm{ mW}$, with $138\textrm{ MHz}$ red detuning from the D2 cooling transition for the main component, with approximately 20\% of the light as repumper. The diameter is about $15\textrm{ mm}$ at the entrance and slightly converges to match the expanding atomic beam through the slower. The horizontal MOT beams have $40\textrm{ mm}$  diameter and the vertical beams have $20\textrm{ mm}$ diameter. During loading, the cooling light is $43\textrm{ MHz}$ red detuned from the D2 cooling transition ($F = 3/2 \to F'$) with $20\textrm{ mW}$ in each beam, and the repumping beams are $38\textrm{ MHz}$ red detuned from the repumping transition ($F = 1/2 \to F'$) with a power of $4\textrm{ mW}$ in each beam.  We at best have $1.2\times10^9$ atoms in the MOT after $10\textrm{ s}$ loading.
We compress the MOT (known as CMOT) by increasing the gradient from $160\textrm{ }\mu\textrm{T}/\textrm{mm}$ to $480\textrm{ }\mu\textrm{T}/\textrm{mm}$, simultaneously shifting the cooling and repumping light $20\textrm{ MHz}$ closer to the transition frequency, while the cooling and repumping powers decrease to 9.1\% and 3.2\% of their previous values, respectively.

\begin{figure}
\includegraphics[width=0.5\textwidth]{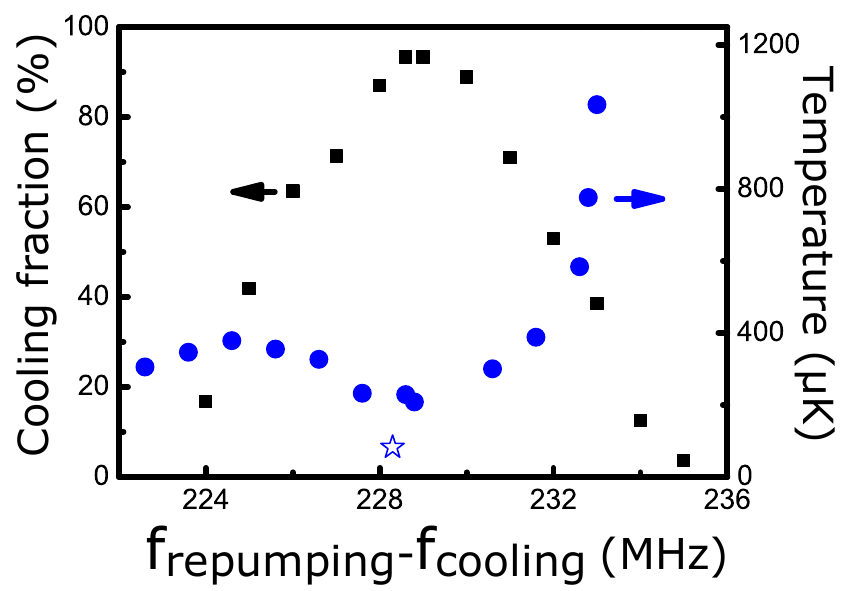}
\caption{\label{fig:D1cooling} Cooling fraction (black squares) and temperature (blue circles) after two stage D1 molasses cooling without canceling stray field (we estimate the field to be $300\textrm{ mG}$). The cooling is only a small improvement, but the frequency sensitivity shows that the gray molasses cooling mechanism is working. After canceling the residual field to better than $10\textrm{ mG}$, the temperature dropped dramatically to $80\textrm{ }\mu\textrm{K}$ (blue star)}
\end{figure}

\section{Gray molasses}
To get a better initial Phase Space Density (PSD) for evaporation, we use D1 gray molasses cooling before loading atoms into our dipole trap. The D1 gray molasses cooling is a very effective sub-Doppler cooling method for lithium that relies on interference between two components of light with a relative frequency difference matched to the ground state hyperfine levels.\cite{0295-5075-100-6-63001,PhysRevA.87.063411,Burchianti:2014fr,Sievers:2015fk}. Our molasses beams share windows with the MOT beams but have small beam diameters of $11\textrm{ mm}$ for the horizontal direction and $8\textrm{ mm}$ for the vertical direction. The total power is $137\textrm{ mW}$ with a fraction of 5\% of repumper. In order to optimize the captured fraction and the final temperature, we employ a two-stage procedure. The first stage begins after quickly switching off the CMOT gradient and D2 cooling beams, and lasts about $5\textrm{ ms}$. The total D1 light intensity is $400 \textrm{ mW}/\textrm{cm}^2$. The second stage lasts only $\textrm{0.2 ms}$, during which the AOM is ramped down to produce 25\% of the starting intensity. The effectiveness of D1 gray molasses is quite sensitive to the two-photon detuning, which is precisely controlled by an AOM. We plot the temperature against two photon detuning in Figure \ref{fig:D1cooling}. The D1 gray molasses, like other sub-Doppler schemes, is sensitive to residual magnetic fields, so we zero the main coil current and use the shim coils to cancel the residual field. Although careful field cancellation was necessary to achieve significant temperature reduction, all of the other conditions could be optimized under larger stray field conditions. The residual field cancellation was adjusted and verified by measuring radiofrequency transitions between hyperfine states. Somewhat surprisingly given the extreme sensitivity, we were not successful in using the cooling efficacy alone to find the zero field condition.  Under optimal conditions, we have close to 100\% of the atoms cooled and the temperature reduced to $80\textrm{ }\mu\textrm{K}$. Since the single photon detuning isn’t critical to the D1 gray molasses, the light shift from the dipole laser is not harmful and we increase the optical dipole power during the D1 molasses. At the end of D1 molasses cooling, we shut off the repumper $0.1 \textrm{ ms}$ before the cooling beam, so the atoms are optically pumped into the lowest two hyperfine states.

\section{Evaporation and BEC}
In what follows we label the absolute lowest three hyperfine levels as $|1\rangle$, $|2\rangle$, and $|3\rangle$, in increasing order of energy. These are adiabatically connected to the levels $\left|m_s,m_I\right\rangle = \left|-1/2,1\right\rangle$, $\left|-1/2,0\right\rangle$, and $\left|-1/2,-1\right\rangle$, respectively, at high field. The evaporation starts with loading a $|1\rangle$ and $2\rangle$ mixture into a crossed optical dipole trap. This hyperfine population results from turning off the component of light near the $F = 1/2$ transition at the end of the molasses phase. The dipole trap is formed by two beams of $1070\textrm{ nm}$ light intersecting at $13^\circ$ on the atomic cloud. Each beam has a maximum $65\textrm{ W}$ power with a waist of $32\textrm{ um}$. The power is controlled by an AOM, to which we apply a $1\textrm{ MHz}$ frequency modulation to extend the trap volume at the early stages of evaporation. The laser power ramps up in $2\textrm{ ms}$ to its maximum during the CMOT phase. The magnetic field ramps up to $79.0\textrm{ mT}$ in $60\textrm{ ms}$ after D1 gray molasses cooling. The laser power remains constant $10\textrm{ ms}$ after the field ramp up finishes.  At this point about $2.5\times10^6$ atoms of each spin state are loaded into the dipole trap and the experiment is ready for forced evaporation. 

\begin{figure}
\includegraphics[width=0.5\textwidth]{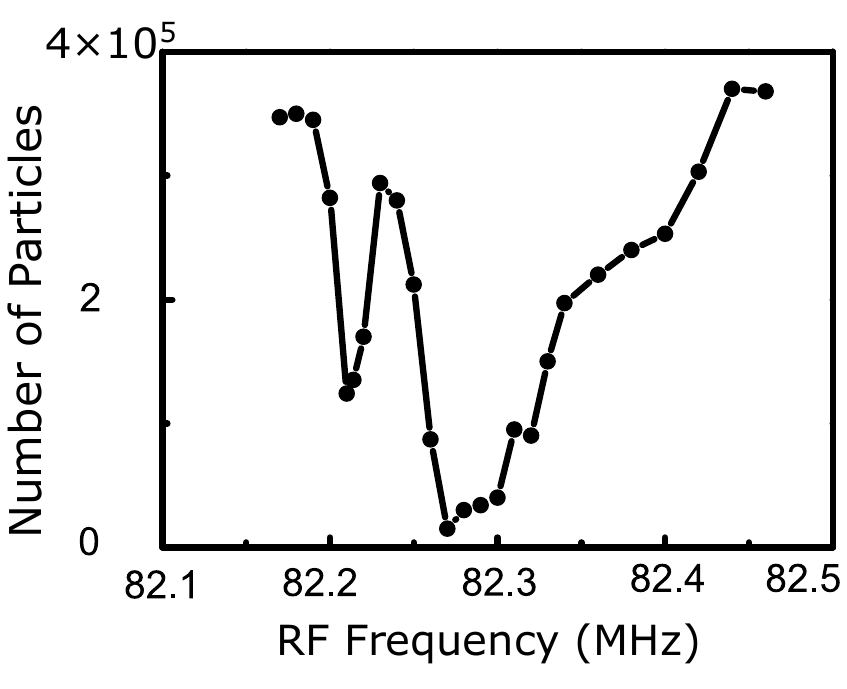}
\caption{\label{fig:moleculepeak} RF spectroscopy. The left dip is the free-to-free resonance of the $|2\rangle$ to $|3\rangle$ transition, while the right dip is the bound-to-free transition indicating the existence of Feshbach molecules. The frequency difference between these two dips is about $60(10)\textrm{ kHz}$, close to the binding energy of $2.2\textrm{ }\mu\textrm{K}$ by calculation.}
\end{figure}

\begin{figure}
\includegraphics[width=0.5\textwidth]{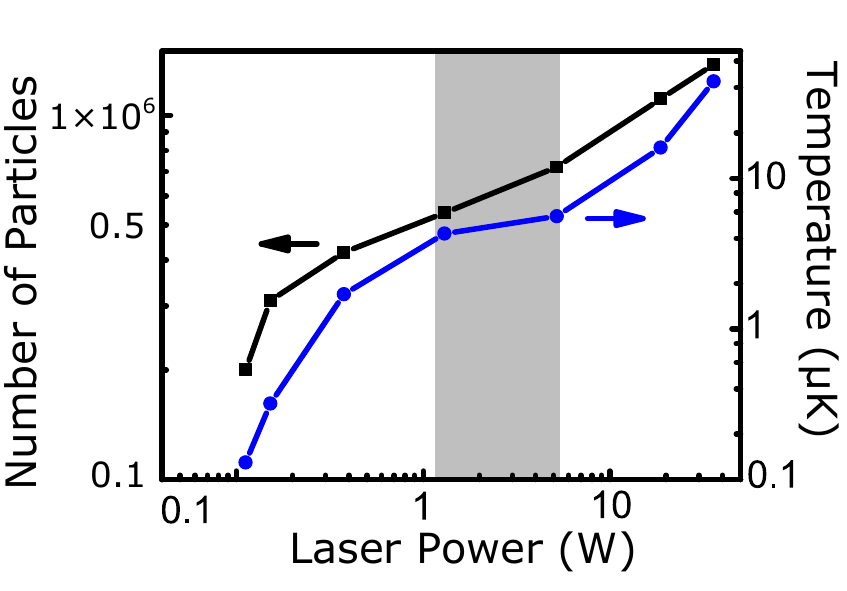}
\caption{\label{fig:EVPspin12} The atom number and temperature during evaporation for $|1\rangle-|2\rangle$ molecular BEC. The gray region indicates where we compress the dipole trap. }
\end{figure}

We have made molecular Bose-Einstein condensates in both $|1\rangle-|2\rangle$ and $|2\rangle-|3\rangle$ mixtures. In order to make a $|1\rangle-|2\rangle$ molecular BEC, we use forced evaporation. The first stage of evaporation lasts $700\textrm{ ms}$, during which we ramp down the laser power to decrease the temperature, and the magnetic field shifts to $75.1\textrm{ mT}$ for a convenient binding energy to generate Feshbach molecules. We confirm the existence of molecules by radiofrequency spectroscopy between the $|2\rangle$ and $|3\rangle$ states as shown in Fig. \ref{fig:moleculepeak}. The binding energy can be estimated by the formula $E_B/k_B=\hbar^2/(ma^2k_B)=2.2\textrm{ }\mu\textrm{K}$, where $E_B$ is the binding energy and $a=188\textrm{ nm}$ at $75.1\textrm{ mT}$ is the s-wave scattering length\cite{RevModPhys.82.1225}. We then spend $600\textrm{ ms}$ to gradually decrease the modulation and increase the atom density, and then a further $800\textrm{ ms}$ of evaporation before the molecular BEC emerges. The atom number and temperature during evaporation are plotted in Fig. \ref{fig:EVPspin12}.

\begin{figure}
\includegraphics[width=0.5\textwidth]{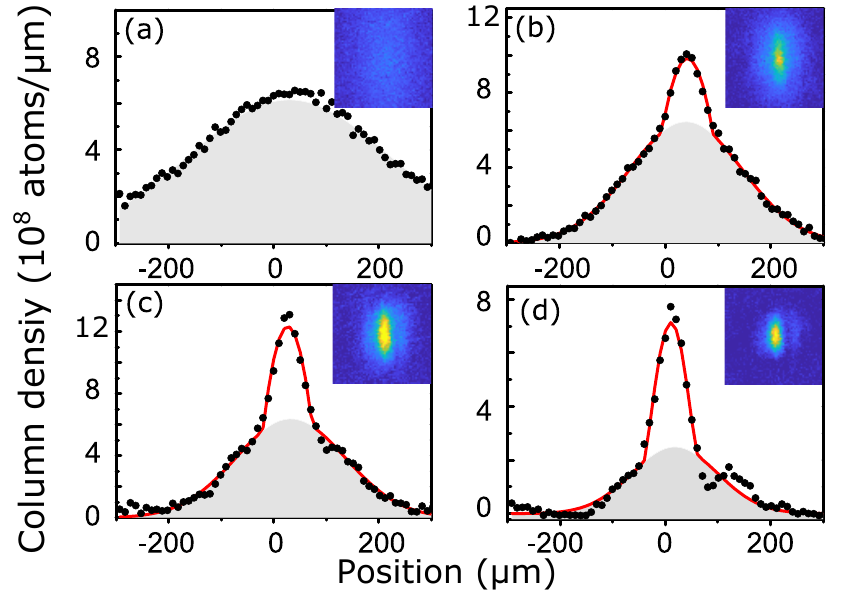}
\caption{\label{fig:Spin12BEC} Bimodel fitting of column density and false color optical density profile after $11\textrm{ ms}$ ballistic expansion when evaporating a $|1\rangle-|2\rangle$ mixture to successively lower dipole laser powers ($P$). (a) $P = 247\textrm{ mW}$, where temperature is $T = 1.9\textrm{ }\mu\textrm{K}$, and molecule number $N = 2.7\times10^5$. No condensation can be observed. (b) $P = 139\textrm{ mW}$, $T = 0.31\textrm{ }\mu\textrm{K}$, and $N = 1.9\times10^5$. $10\%$ of molecules are condensed. (c) $P = 88\textrm{ mW}$, where $T = 0.06\textrm{ }\mu\textrm{K}$, and $N = 1.6\times10^5$. $19\%$ of molecules are condensed. (d) $P = 57\textrm{ mW}$, where $T = 0.05\textrm{ }\mu\textrm{K}$, and $N = 9\times10^4$. $36\%$ atoms are condensed. The ideal gas Fermi temperature and BEC critical temperature would be $0.57\textrm{ }\mu\textrm{K}$ and $0.30\textrm{ }\mu\textrm{K}$, respectively, under these conditions. The fits are done using a 2D Gaussian profile plus a truncated elliptic paraboloid.}
\end{figure}

The bimodal profile is one of the easiest tests for the existence of BEC\cite{Zwierlein:2003ch}. However, for ${}^6\textrm{Li}$ in the strongly interacting region, a special imaging procedure is used to make the profile more obvious\cite{M_Kohnen_Thesis_2008}. We first ramp down the field from $79.0\textrm{ mT}$ to $50.0\textrm{ mT}$ in $1\textrm{ ms}$ at the start of the time-of-flight, so the molecules can expand ballistically. At this field the binding energy is too deep to image, so we then ramp up to $85.0\textrm{ mT}$ to unbind the molecules during the second half of the time-of-flight period. This increasing field ramp takes about $5\textrm{ ms}$, limited by the coil inductance. Our high field experiments are greatly aided by the beat lock of the D2 laser used for imaging, which allows absorption images at GHz Zeeman shifts. Using this procedure, a clear signature of molecular BEC can be seen at the end of the evaporation (see Fig. \ref{fig:Spin12BEC}). We estimate that we can have condensates with nearly $10^5$ molecules, about $36\pm4$\% of which are in the BEC. The total number is determined by standard absorption imaging methods, and a bimodal fit is used to estimate condensation fraction. There is some systematic uncertainty to this estimate due to the rapidly changing scattering length during the initial time-of-flight period, which prevents true ballistic expansion. Since these effects cover $1\textrm{ ms}$ out of the $11\textrm{ ms}$ time of flight period, we estimate this uncertainty at $10\%$, comparable to imaging shot noise uncertainty. Similar considerations apply to the temperature, which is estimated to be around $0.05\textrm{ }\mu\textrm{K}$, based on the thermal portion of the image. At lower laser powers, the temperature does not decrease significantly, suggesting there may be competing heating processes\cite{Deng:2015cu}.

\begin{figure}
\includegraphics[width=0.5\textwidth]{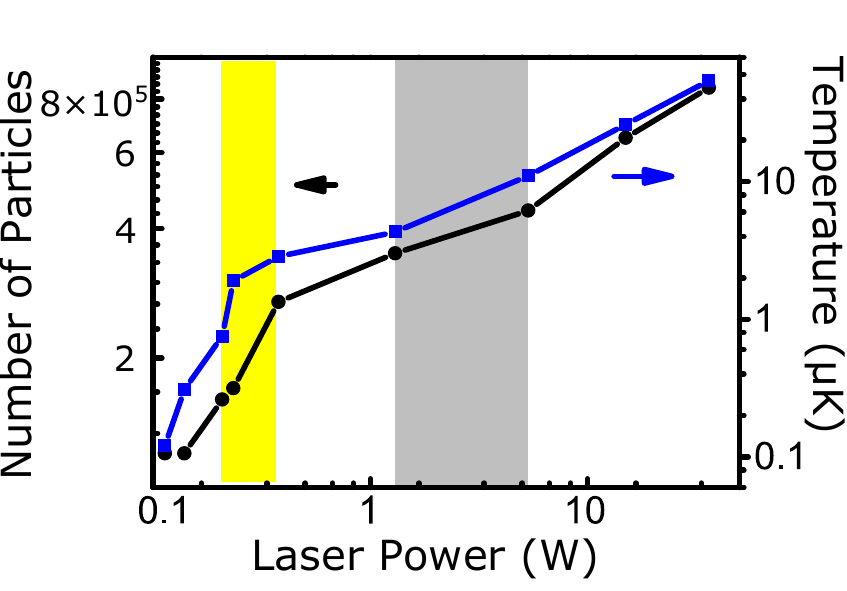}
\caption{\label{fig:EVPspin23} The atom number and temperature during evaporation for $|2\rangle-|3\rangle$ molecular BEC. The gray region indicate we compress the volume of dipole trap.we ramp the magnetic field to increase the binding energy when crossing the yellow region}
\end{figure}

\begin{figure}
\includegraphics[width=0.5\textwidth]{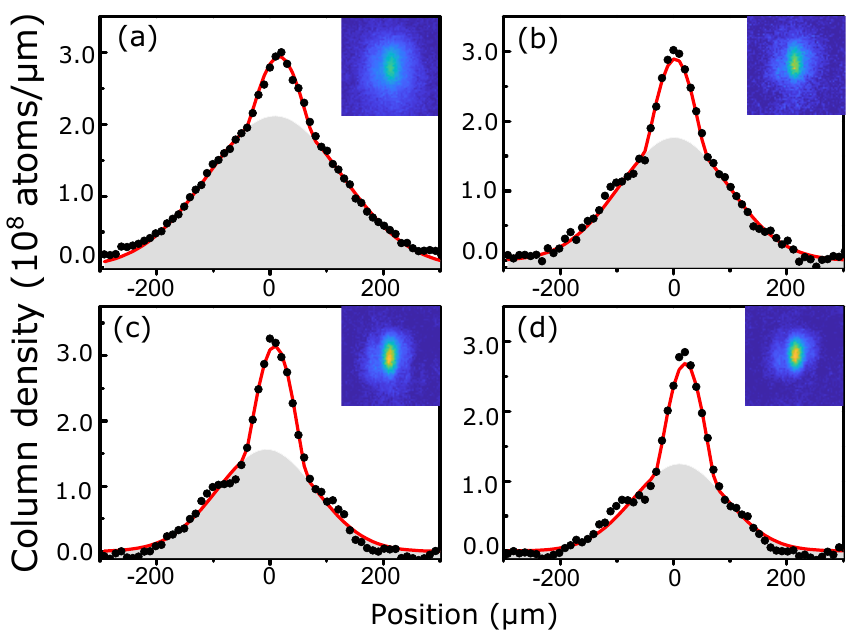}
\caption{\label{fig:Spin23BEC} Bimodal fitting of column density and false color optical density profile after $11\textrm{ ms}$ ballistic expansion when evaporating a $|2\rangle-|3\rangle$ mixture to successively lower dipole laser powers ($P$). (a) $P = 112\textrm{ mW}$, where temperature $T = 0.13\textrm{ }\mu\textrm{K}$, and molecule number $N = 8.1\times10^4$. $9\%$ of molecules are condensed. (b) $P = 104\textrm{ mW}$, where $T = 0.12\textrm{ }\mu\textrm{K}$, and $N = 6.1\times10^4$. $16\%$ of molecules are condensed. (c) $P = 88\textrm{ mW}$, where $T = 0.06\textrm{ }\mu\textrm{K}$, and $N = 4.8\times10^4$. $23\%$ of molecules are condensed. (d) $P = 68\textrm{ mW}$, where $T = 0.05\textrm{ }\mu\textrm{K}$, and $N = 3.2\times10^4$. $28\%$ of molecules are condensed. The ideal gas Fermi temperature and BEC critical temperature would be $0.44\textrm{ }\mu\textrm{K}$ and $0.23\textrm{ }\mu\textrm{K}$, respectively, under these conditions. The fits are done using a 2D Gaussian profile plus a truncated elliptic paraboloid. Each image shown is the average of 5 acquisitions.}
\end{figure}

To make a $|2\rangle-|3\rangle$ molecular BEC, we start by loading a $|1\rangle-|2\rangle$ mixture into crossed dipole trap. Then, we transfer atoms to the desired spin states, which requires a two-pulse sequence: $|2\rangle \to |3\rangle$, followed by $|1\rangle \to |2\rangle$. A challenge to this process are collisions and field inhomogeneity, as both destroy the coherence of the transfer. To avoid collisions, one might like to work around $55.0\textrm{ mT}$ where scattering cross sections are small. However, at this field the difference of magnetic moment is much larger, and our inhomogeneous field makes such transfers poor. Instead we opt to do fast transfers at $73.6\textrm{ mT}$ and accept the collisional incoherence. We apply a $0.36\textrm{ ms}$ $\pi$ pulse to transfer the atoms from $|2\rangle$ to $|3\rangle$ and then a $0.48\textrm{ ms}$ $\pi$ pulse to transfer the atoms from $|1\rangle$ to $|2\rangle$. About $50\%$ of the atoms can be successfully driven to $|3\rangle$ by the first pulse. We believe the efficiency is limited by the decoherence due to the existence of $|1\rangle$. The transfer efficiency of our $\pi$ pulse can be as high as 80\% when no $|1\rangle$ atoms exist, and Landau-Zener sweeps have similar performance. The imperfect transfer here is most likely due to the curvature of the magnetic field. Currently we limit the RF power to $2\textrm{ W}$, so coherence may improve with faster transfers and shorter pulse times. A light pulse resonant with state $|1\rangle$ clears the residual atoms right after the second RF pulse to avoid depletion due to inelastic collisions. We have about $1\times10^6$ $|2\rangle$ atoms and $8\times10^5$ $|3\rangle$ atoms remaining in the trap with a temperature of $54\textrm{ }\mu\textrm{K}$. We evaporate the $|2\rangle-|3\rangle$ mixture at $75.7\textrm{ mT}$ to $11\textrm{ }\mu\textrm{K}$ in $200\textrm{ ms}$. Then we decrease the modulation of the AOM to compress the dipole trap within $400\textrm{ ms}$. After that during the next $600\textrm{ ms}$ of evaporation we ramp the field to $74.2\textrm{ mT}$ where we expected the $|2\rangle-|3\rangle$ binding energy is similar to $|1\rangle-|2\rangle$ mixture at $75.1\textrm{ mT}$. The binding energy is $80\textrm{ kHz}$ as measured with the same RF transfer method used for the $|1\rangle-|2\rangle$ mixture. We observe the BEC by applying a further $700\textrm{ ms}$ evaporation. The atom number and temperature during evaporation are shown in Fig. \ref{fig:EVPspin23}. We use the same imaging technique to identify the bimodal distribution and determine the BEC condensate fraction as shown in Fig. \ref{fig:Spin23BEC}. At best we can have $28\pm3$\% atoms condensed with a total number of $3.2\times10^4$. Although the RF transfer incurs a significant atom number loss, we can see that it does not prevent the formation of condensates. 

\section{Conclusion}
In conclusion, we have demonstrated an all-optical method for producing molecular BECs of ${}^6\textrm{Li}$ in either the $|1\rangle-|2\rangle$ or $|2\rangle-|3\rangle$ hyperfine mixtures. In principle it would be easy to extend this to the $|1\rangle-|3\rangle$ mixture as well. These experiments pave the way for future investigations of degenerate Fermi gases and detailed comparisons of many-body properties between mixtures with identical two-body collisional properties.

\begin{acknowledgments}
We acknowledge support from the Air Force Office of Scientific Research, Young Investigator Program, through grant number FA9550-18-1-0047.
\end{acknowledgments}

\bibliography{cvp,cvppapers}

\end{document}